\documentclass[preprint,showpacs,preprintnumbers,amsmath,amssymb]{revtex4}

\usepackage{dcolumn}
\usepackage{bm}

\begin{document}


\title{Geodesics, Mass and the Uncertainty Principle in a Warped de Sitter Space-Time}

\author{Jose A. Magpantay}
\email{jose.magpantay@up.edu.ph}
\affiliation{National Institute of Physics and Technology Management Center, University of the Philippines, Quezon City, Philippines\\}

\date{\today}

\begin{abstract}
We present the explicit solution to the geodesic equations in a warped de Sitter space-time proposed by Randall-Sundrum. We find that a test particle moves in the bulk and is not restricted on a 3-brane (to be taken as our universe). On the 3-brane, the test particle moves with uniform velocity, giving the appearance that it is not subject to a force. But computing the particle's energy using the energy-momentum tensor yields a time-dependent energy that suggests a time-dependent mass. Thus, the extra force, which is the effect of the warped extra dimension on the particle's motion on the 3-brane, does not change the velocity but the mass of the particle. The particle's motion in the bulk also results in a time-dependent modification of the Heisenberg uncertainty principle as viewed on the 3-brane. These two results show that the classical physics along the extra dimension results in the time-dependence of particle masses and the uncertainty principle. If the particle masses are time-independent and Heisenberg's uncertainty principle is to remain unchanged, then there must be a non-gravitational force that will restrict all particles on the 3-brane. Finally, we just note that although classically, these time-dependent corrections on the 3-brane can be removed, quantum mechanical corrections along the extra dimension will restore back the problem. 
\end{abstract}

\maketitle

1.0 The Randall-Sundrum models (RS1 \cite{Randall1} and RS2 \cite{Randall2}) were proposed as possible explanation to the difference in scales between gravity and the electroweak physics (one of the hierarchy problems). In contrast to the Arkani-Hamed, Dimopoulos and Dvali (ADD) \cite{Arkani-Hamed} solution to the same problem where the extra dimensions are flat and range in size from $10^{12}$ m (one extra dimension) to $\leq 10^{-12}$ m (four or more extra dimensions), in RS2 the extent of the lone extra dimension is $\infty$ and it is the warping factor that accounts for the difference in strength between the electroweak force and gravity.

A basic assumption in the Randall-Sundrum model (and also in ADD) is that all particles, except the graviton, are constrained on the 3-brane, which is identified as our universe, by some unknown string mechanism. In this paper, we would investigate the motion of a particle if the string constraint is removed. In effect, we would derive the geodesics of a particle in a warped de Sitter space-time. We will also compute the particle's energy from the energy-momentum tensor and show that it is time-dependent, suggesting a time-dependent mass. 

The explicit geodesic solutions show that a test particle with mass will move in the bulk. This is consistent with the previous works of a number of authors \cite{Liu}, \cite{Youm}, \cite{Ponce de Leon1}. In particular, Liu and Peng showed that in RS1, a particle between the two boundary branes will move in the bulk toward the visible brane resulting in instability thus necessitating the introduction of a fluid in the bulk to restore stability. What is new in our computation is the result that on the 3-brane the particle will exhibit a uniform velocity motion thus suggesting that it is not subject to a force. However, this is false as the energy computation will show.

Starting from the geodesic solutions, we use the energy-momentum tensor to compute the particle's energy. We give an explicit expression for the particle's energy and show that it is time-dependent. Furthermore, the time-dependent energy points to a time-dependent mass. This means the particle on the 3-brane is subject to an extra force due to the presence of extra coordinates, in this case a warped extra dimension. This result has already been known in the works of other people \cite{Youm}, \cite{Wesson1}, \cite{Ponce de Leon2}. What is new in our derivation is the use of the energy-momentum tensor in deriving the result. Also, the energy that was derived in this paper is clearly the extension of the Einstein energy-mass relation if motion in the warped extra dimension is allowed. 

We will also explore the consequence of the warped space-time on the uncertainty principle. If the particle is allowed motion in the bulk, we show that the Heisenberg uncertainty principle will be modified. The modification is not a quantum gravity effect (it is not Planck length dependent) but a simple kinematical effect of motion in the bulk on the 3-brane dynamics.

We note that if we begin with the fact that the measured masses of particles are constants and the Heisenberg uncertainty principle is well tested at the present energy scales, then we must posit the existence of a non-gravitational extra force, sometimes attributed to string physics, to restrict the particles on the 3-brane. We give an expression for this extra force. This non-gravitational force is known in the literature.       

Finally, we point out the effect of the quantum dynamics of the particle in the bulk. We argue that even with the extra non-gravitational force that restricts the particle on the 3-brane, the quantum effects will make the particle, at the very least fluctuate, along the warped extra dimension. This will trigger small changes on the 3-brane, which must cause small corrections to the masses of the particles and to the Heisenberg uncertainty principle on the 3-brane.
 
2.0 The measure in the warped de Sitter space-time used by Randall and Sundrum is given by    
\begin{equation}\label{sha}
\ d\tau^2 = \exp{(-k\lvert y \rvert)}\left[(dx^0)^2-d\vec{x}\cdot d\vec{x}\right]-dy^2,
\end{equation}
where y is the extra dimension and the warping strength k is determined from the cosmological constant and the fundamental mass scale in the bulk. The non-vanishing Christoffel symbols are
\begin{subequations}\label{sha2}
\begin{align}
\ \Gamma^0_{04} = \Gamma^0_{40} = \Gamma^i_{4i} = \Gamma^i_{i4} = \mp(\frac{1}{2})k,\\
\ \Gamma^4_{00} = -\Gamma^4_{ii} = \mp(\frac{1}{2})k\exp{(-k\lvert y \rvert)},
\end{align}
\end{subequations}
where the upper/lower signs correspond to $y>0/y<0$ and is a consequence of $\dfrac{d\lvert y \rvert}{dy} = sign(y)$. The geodesic equations are
\begin{subequations}\label{sha3}
\begin{gather}
\dfrac{d^2x^0}{d\tau^2} \mp k\dfrac{dx^0}{d\tau}\dfrac{dy}{d\tau}=0,\\
\dfrac{d^2x^i}{d\tau^2} \mp k\dfrac{dx^i}{d\tau}\dfrac{dy}{d\tau}=0,\\
\dfrac{d^2y}{d\tau^2} \pm \frac{1}{2}k\exp{(-k\lvert y \rvert)} \left[\dfrac{dx^i}{d\tau}\dfrac{dx^i}{d\tau}-\dfrac{dx^0}{d\tau}\dfrac{dx^0}{d\tau}\right]=0.
\end{gather}
\end{subequations}
These are non-linear, ordinary differential equations, which are exactly solvable as we will show below.\\
Defining $u^\mu=\dfrac{dx^\mu}{d\tau}$, for $\mu=0,1,2,3$, we find that equations \eqref{sha3} are solved by
\begin{equation}\label{sha4}
u^\mu=\beta^\mu\exp{(k\lvert y \rvert)},
\end{equation}
where $\beta^\mu=\dfrac{dx^\mu}{d\tau}(y=0)$. Using \eqref{sha} in equation (3c), we find
\begin{equation}\label{sha5}
\dfrac{d^2y}{d\tau^2} \mp\frac{1}{2}k\left[1+(\dfrac{dy}{d\tau})^2\right]=0.
\end{equation}
This equation shows that $y=0$ is an unstable point because displacing the particle to the right/left results in a force in the same direction which will move the particle farther away. This is the rationale for introducing a fluid in the bulk in the paper of Liu and Peng (see reference 4).\\
Integrating equation (5) yields
\begin{equation}\label{sha6}
\dfrac{dy}{d\tau}=\tan{\left[\arctan{v_0} \pm\frac{1}{2}k\tau\right]},
\end{equation}
where $v_0=\dfrac{dy}{d\tau}(\tau=0)$. Integrating equation (6) gives
\begin{equation}\label{sha7}
y(\tau)=\mp\frac{2}{k}\ln{\left[\cos{(\frac{1}{2}k\tau)}\mp v_0\sin{(\frac{1}{2}k\tau)}\right]}.
\end{equation}
Note that the solution satisfies $y(\tau=0)=0$, which means $\beta^\mu=\dfrac{dx^\mu}{d\tau}(y=0)=\dfrac{dx^\mu}{d\tau}(\tau=0)$, i.e., the constants $\beta^\mu$ are also initial conditions and not only boundary conditions. We can also rewrite equation (6) as
\begin{equation}\label{sha8}
\dfrac{dy}{d\tau}=\pm\left[(1+v^2_0)\exp{(k\lvert y \rvert)}-1\right]^{\frac{1}{2}},
\end{equation}
which clearly satisfies $\dfrac{dy}{d\tau}(\tau=0)=\dfrac{dy}{d\tau}(y=0)=v_0$ and is consistent with the reading of equation (5). Substituting equation (7) in equation (4), we find
\begin{equation}\label{sha9}
x^\mu(\tau)=b^\mu+\beta^\mu\left[\frac{1}{2}\tau(1+v^2_0)+(1-v^2_0)\frac{1}{2k}\sin{(k\tau)}\mp\frac{4}{k}\sin^2{(\frac{1}{2}k\tau)}\right],
\end{equation}
where $b^\mu=x^\mu(\tau=0)$. We will take $b^0=0$ so that the time $x^0$ on the 3-brane is initially synchronized with $\tau$. Finally, we relate the constants of integration by taking note of equation (1) at $y=0$, yielding
\begin{equation}\label{sha10}
1=(\beta^0)^2-\vec{\beta}\cdot\vec{\beta}-v^2_0.
\end{equation} 
Equations (7) and (9) give the solutions to the geodesic equations for a test particle with mass in a Randall-Sundrum space-time. It shows the particle moving in the bulk. Time ($x^0$) dependence of $\vec{x}$ and y can be shown by solving for $\tau=\tau(x^0)$ in the $\mu=0$ component of equation (9). The dynamics on the 3-brane is simply given by
\begin{equation}\label{sha11}
\vec{x}=\vec{b}+\dfrac{\vec{\beta}}{\beta^0}x^0,
\end{equation}
which is a uniform velocity motion hinting of free particle dynamics. The y dynamics as a function of $x^0$ is more involved because solving for $\tau(x^0)$ can only be done by solving a polynomial equation, which results from expanding the sine functions in equation (9) in a Maclaurin series. The fact that $y(x^0)$ results in a time-dependent energy for the particle means that as viewed on the 3-brane, the particle is subject to a force that does not change its velocity ($\dfrac{d\vec{x}}{dx^0}=\dfrac{\vec{\beta}}{\beta^0} = constant$) but only changes its energy. As we will show in the next section, the energy is written in a form which hints of time-dependent mass.

	3.0 In a general n dimensional space-time, the energy-momentum tensor of a particle is given by \cite{Weinberg}
\begin{equation}\label{sha12}
T^{ab}=M(-g)^{-\frac{1}{2}}\int^{\infty}_{-\infty}d\tau\dfrac{dx^a}{d\tau}\dfrac{dx^b}{d\tau}\delta^n(x-\tilde{x}(\tau)),
\end{equation}
where the tilde quantities are solutions of the geodesic equation and M is the particle's invariant mass. The particle's energy is given by
\begin{equation}\label{sha13}
\epsilon=\int T^{00}d^{n-1}x,
\end{equation}
and evaluating this for the warped de Sitter space-time yields
\begin{equation}\label{sha14}
\epsilon= \dfrac{M\exp{(\frac{5}{2}k\lvert \tilde{y} \rvert)}}{(1-\dfrac{\vec{\beta}\cdot\vec{\beta}}{(\beta^0)^2})^{\frac{1}{2}}}
\left[1-\exp{(k\lvert \tilde{y} \rvert)}(\dfrac{d\tilde{y}}{dx^0})^2(1-\dfrac{\vec{\beta}\cdot\vec{\beta}}{(\beta^0)^2})^{-1}\right]^{-\frac{1}{2}},
\end{equation}
where $\tilde{y}$ is given by \eqref{sha7} expressed in terms of $\tau=\tau(x^0)$, which is to be solved from the $\mu=0$ component of \eqref{sha9}. Equation \eqref{sha14} also takes into account $\dfrac{d\vec{x}}{dx^0}=\dfrac{\vec{\beta}}{\beta^0}$ and clearly implies that the time-dependent energy is due to a time-dependent mass. It also clearly suggests that this is the modification of Einstein's $\epsilon=\dfrac{m}{\sqrt{1-\frac{v^2}{c^2}}}c^2$ if motion in the bulk is taken into account. In the limit of fixed $\tilde{y}$, equation \eqref{sha14} is just the special relativistic energy of a free particle with time-independent rest mass $m_0$ given by $m_0=M\exp{(\frac{5}{2}k\left|\tilde{y}\right|)}$.

Since the particle moves in the bulk as given in equation \eqref{sha7}, the rest mass energy as viewed on the 3-brane is time-dependent and is given by ($\vec{\beta}=0$)
\begin{equation}\label{sha15}
m(x^0)=M\exp{(\frac{5}{2}k\left|\tilde{y}\right|)}\left[1-\exp{(k\left|\tilde{y}\right|)}(\dfrac{d\tilde{y}}{dx^0})^2\right]^{-\frac{1}{2}}.
\end{equation}

This is similar to Ponce de Leon's equation (13) \cite{Ponce de Leon2}, which he derived using the Hamilton-Jacobi equation in 5D.
 
	4.0 So far, we explored the consequences of the motion in the bulk in warped extra dimension on the particle's kinematics and mass as viewed on the 3-brane. We will now explore the consequence on non-relativistic quantum theory. We do this in terms of the effect on the uncertainty principle. The momentum of the particle on the 3-brane is $p_i=mv_i$, where the mass is time-dependent as given by equation (15). The uncertainty in the momentum measurement is given by
\begin{equation}\label{sha16}
\Delta p_i=m\Delta v_i+\Delta mv_i,
\end{equation}
where the first term is the usual uncertainty in momentum and thus equal to $\frac{\hbar}{\Delta x_i}$. The second term is the uncertainty in the mass because during the measurement with time resolution $\delta x^0$ the particle's mass changes because of the particle's motion in the bulk. This uncertainty in mass is given by
\begin{equation}\label{sha17}
\Delta m=\frac{dm}{dy}\frac{dy}{dx_0}\delta x^0,
\end{equation}
where the derivatives are derived from equations (15), (8) and (9). But $\delta x^0v_i=\Delta x_i$, i.e., the uncertainty in the x measurement is equal to the time resolution multiplied by the particles velocity. Thus, we find
\begin{equation}\label{sha18}
\Delta p_i=\frac{\hbar}{\Delta x_i}+\frac{dm}{dy}\frac{dy}{dx^0}\Delta x_i.
\end{equation}
Note, this form of the uncertainty principle is reminiscent of the dual generalized uncertainty principle for de Sitter gravity \cite{Magpantay}. There, it is the cosmological constant, which drives the accelerated expansion of the universe that provides the small correction to the Heisenberg uncertainty principle. Here, it is the motion in the bulk that adds uncertainty in the measurement in the particle's momentum through the uncertainty in the particle's mass during the measurement process. The correction to the Heisenberg uncertainty principle is time-dependent.

We note that in an earlier paper, Wesson \cite{Wesson2} suggested that the uncertainty relation in 4D must be a consequence of classical certainty in 5D. He derived a relation that shows the 'fractional change in the position of the particle in the fifth dimension has to be matched by a change of its dynamical quantities in 4D spacetime'. Our derivation is more straightforward and shows not only the Heisenberg uncertainty principle but also how it is corrected by the time-dependent mass coming from the motion of the particle along the warped extra dimension.  

	5.0 The motion in the bulk led to two interesting corrections to present day physics. These are the time-dependent mass of the particles and time-dependent correction to the Heisenberg uncertainty principle. Either these corrections are very small (Wesson \cite{Wesson1} estimated the time rate of change of mass $\frac{1}{m}\frac{dm}{dt}\approx 10^{-18} sec^{-1}$, which is the inverse of the age of the universe) or not present at all (the Heisenberg principle is an established pillar of quantum mechanics and correcting it in a time-dependent way does not seem to be 'natural' for a fundamental law). If the corrections are removed classically, then there must be an extra force that should restrict the particles on the 3-brane. This extra non-gravitational force is known (see for example equation (26) of \cite{Ponce de Leon2}). Unfortunately, although the particle can be restricted classically on the 3-brane, quantum mechanics along the extra dimension will induce fluctuations along this direction. We know this to be true from the hydrogen atom. Classically, a central force leads to a motion on a plane yet quantum mechanically the electron cloud shows the electron leaking out to the third dimension. The same thing must be true with dynamics in the bulk - i.e., quantum mechanics will make the particle 'leak out' to the bulk. What happens now to the particle masses and the Heisenberg uncertainty principle on the 3-brane? For these issues to be resolved, a full quantum treatment in the bulk should be made. 
        
\begin{acknowledgments}
This research was supported in part by the Research and Creative Work Grant of the University of the Philippines.
\end{acknowledgments}

\end{document}